\documentclass[a4paper,11pt]{article}
\pdfoutput=1 

\usepackage{jinstpub} 

\title{An open source toolkit for the tracking, termination and recovery of high altitude balloon flights and payloads}

\author[a,1]{Paul Clark,\note{Corresponding author.}}
\author[b]{Marc Funk,}
\author[c]{Benjamin Funk,}
\author[c]{Tobias Funk,}
\author[d]{Richard E. Meadows,}
\author[e]{Anthony M. Brown,}
\author[f]{Lun Li,}
\author[g,e]{Richard J. Massey}
\author[h]{and C. Barth Netterfield}

\affiliation[a]{Centre for Advanced Instrumentation, Durham University,\\NETPark Research Institute, Joseph Swan Road, NETPark, Sedgefield, TS21 3FB, UK}
\affiliation[b]{Lonza AG,\\M{\"u}nchensteinerstrasse 38, CH-4002 Basel, Switzerland}
\affiliation[c]{Space Markt,\\124 route de chene, Geneva, CH 1224, Switzerland}
\affiliation[d]{formerly of University of Bristol,\\Senate House, Tyndall Ave, Bristol BS8 1TH, UK}
\affiliation[e]{Centre for Advanced Instrumentation, Durham University,\\South Road, Durham, DH1 3LE, UK}
\affiliation[f]{Department of Physics, Princeton University,\\Washington Road, Princeton, NJ, USA}
\affiliation[g]{Institute for Computational Cosmology, Durham University,\\South Road, Durham, DH1 3LE, UK}
\affiliation[h]{Department of Physics, University of Toronto,\\60 St. George Street, Toronto, ON, Canada}

\emailAdd{paul.clark@durham.ac.uk}

\abstract{We present an open source toolkit of flight-proven electronic devices which can be used to track, terminate and recover high altitude balloon flights and payloads. Comprising a beacon, pyrotechnic and non-pyrotechnic cut-down devices plus associated software, the toolkit can be used to: (i) track the location of a flight via Iridium satellite communication; (ii) release lift and/or float balloons manually or at pre-defined altitudes; (iii) locate the payload after descent. The size and mass of the toolkit make it suitable for use on weather or sounding balloon flights. We describe the technology readiness level of the toolkit, based on over 20 successful flights to altitudes of typically 32,000 m.}


\begin{document}
\maketitle
\flushbottom

\section{Introduction}

The retrieval of hardware from a high altitude balloon (HAB) can become important when many balloons are released (each with a significantly expensive payload) or when one balloon acquires enough data that transmitting it to the ground would be impossible (because of either limited bandwidth or cost). In particular, the University of Durham Centre for Advanced Instrumentation (CfAI) and Institute for Computational Cosmology (ICC) are participants in the design and development of the \textit{Superpressure Balloon-borne Imaging Telescope} (``\textsc{SuperBIT}'') ~\cite{d,e}. This is scheduled to fly for up to 100 days on a NASA superpressure balloon in 2020, during which time it will accumulate terabytes of astronomical imaging. Furthermore, as most of the Southern hemisphere flight will be spent over open ocean, successful recovery of the payload is not guaranteed ~\cite{a}. The ability to recover uncompressed data on a solid state disc whenever the balloon passes over land removes a critical mission risk of loss at sea.

We have developed a hardware retrieval system to release itself from the balloon gondola on command, and descend via parachute; the likely landing site having been predicted from wind data prior to release. The essential subsystems of the parachute-based data retrieval system are:
\begin{itemize}
\item An embedded computer which provides File Transfer Protocol (FTP) access to the terabyte solid state storage via Wi-Fi\texttrademark
\item Global Navigation Satellite System (GNSS) receiver
\item Short Burst Data (SBD) transceiver module
\item Low extraction-force power connector
\item Release mechanism
\end{itemize}

The toolkit described herein was developed to allow the GNSS receiver, SBD transceiver and the release mechanism subsystems to be tested independently.
Indeed, it has proved useful in its own right to allow tracking, termination and recovery of payloads flown on weather or sounding balloon flights. The toolkit comprises:
\begin{itemize}
\item Iridium 9603 Beacon (see section~\ref{sec:beacon})
\item Pyrotechnic Cut-Down Device (see section~\ref{pyro_device})
\item Mechanical Cut-Down Device (see section~\ref{sec:mechanical})
\item Associated Python software (see section~\ref{Python})
\end{itemize}
The design of the beacon is currently in its fifth iteration, and more than 20 successful flights (see section~\ref{sec:flighttests}) have allowed the Technology Readiness Level of the toolkit to be established. 

The intent of this paper is twofold: it presents a flight-proven open-source design which can be used as-is; and summarizes the considerations which went into the design to aid anyone wishing to adapt the design to other purposes. For example, aspects of the design have already been used successfully in ocean buoys (personal communication with James D. Whitlock, Dartmouth College, New Hampshire, USA) and to track icebergs (personal communication with Adam Garbo, Water and Ice Research Laboratory, Carleton University, Ottawa, Canada).

As a future enhancement, 
we have also been investigating the feasibility of controlled descent via autonomous glider. This could add hundreds of kilometers of downrange capability, to access better landing sites, for example away from populated areas but near a road. The glider design and control systems were inspired by the original work of Russell G.\ Smith (personal communications regarding Smith's Automated Information Retrieval System (AIRS)). The technical and safety considerations of operating an autonomous glider at altitude are of course not trivial, and being able to simultaneously develop and demonstrate the technology readiness of the glider is challenging. This will be described in a future paper.

\section{Iridium 9603 Beacon} \label{sec:beacon}

\subsection{Requirements}
It is essential to be able to continuously track the location of the payload, for safety and for retrieval once it has returned to the ground. While it is attached to the balloon, a beacon also allows the location of the balloon to be monitored. The primary requirements for such a beacon are:
\begin{itemize}
\item To function reliably in the operating environment (atmospheric temperatures and pressures) encountered during flights to typically 32,000 m Above Sea Level (ASL)
\item To be able to reliably determine the location of the beacon during all parts of the flight and after landing
\item To be able to reliably transfer small amounts of data (location and control messages) to and from the beacon during all parts of the flight and after landing
\item For the size and mass of the beacon to be well within the lift capacity of a typical weather balloon
\item To be able to operate anywhere, including the polar regions
\end{itemize}

\subsection{Operating Conditions}
\begin{figure}[htbp]
\centering 
\includegraphics[width=.75\textwidth]{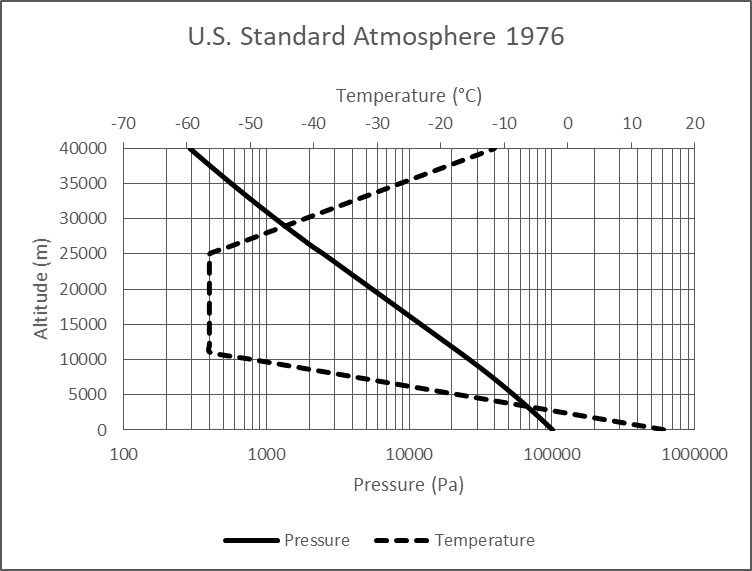}
\caption{\label{fig:Std_Atm} Atmospheric pressure and temperature as defined in the U.S.\ Standard Atmosphere 1976 ~\cite{c}. The temperature between 11,000\,m and 25,000\,m is -56\,\textdegree{}C.}
\end{figure}

The HAB operating environment is challenging, particularly in terms of the temperatures encountered during a flight. Fig.~\ref{fig:Std_Atm} illustrates how atmospheric temperature decreases from +15\,\textdegree{}C at sea level to -56\,\textdegree{}C at 11,000 m. The temperature remains constant until 25,000 m at which point, counter-intuitively, it starts to increase. This temperature profile ignores the effect of solar heating which will increase the temperature of the balloon and payload; the temperature profile can be thought of as the profile encountered during a nighttime flight.

Most commercial electronic components today have a quoted operational temperature limit of -40\,\textdegree{}C. This does not mean that they will immediately stop working at temperatures lower than this, rather that their operation is not guaranteed. The design approach we have taken for the HAB toolkit is to accept that the electronics may not work correctly at some altitudes during the night, but to ensure that it will reset and start operating correctly again after sunrise or upon reaching a warmer altitude. We were able to demonstrate this on the March 2017 superpressure balloon flight where the beacon did indeed stop transmitting overnight and restart the following morning; this is illustrated by the gaps between the groups of waypoints shown in Fig.~\ref{fig:UBSEDS22}. Interestingly the beacon did continue to transmit for the whole of the second night while the balloon was floating in warmer air above Morocco.

\subsection{Design Considerations and Component Selection}
\begin{figure}
\centering 
\includegraphics[width=.75\textwidth]{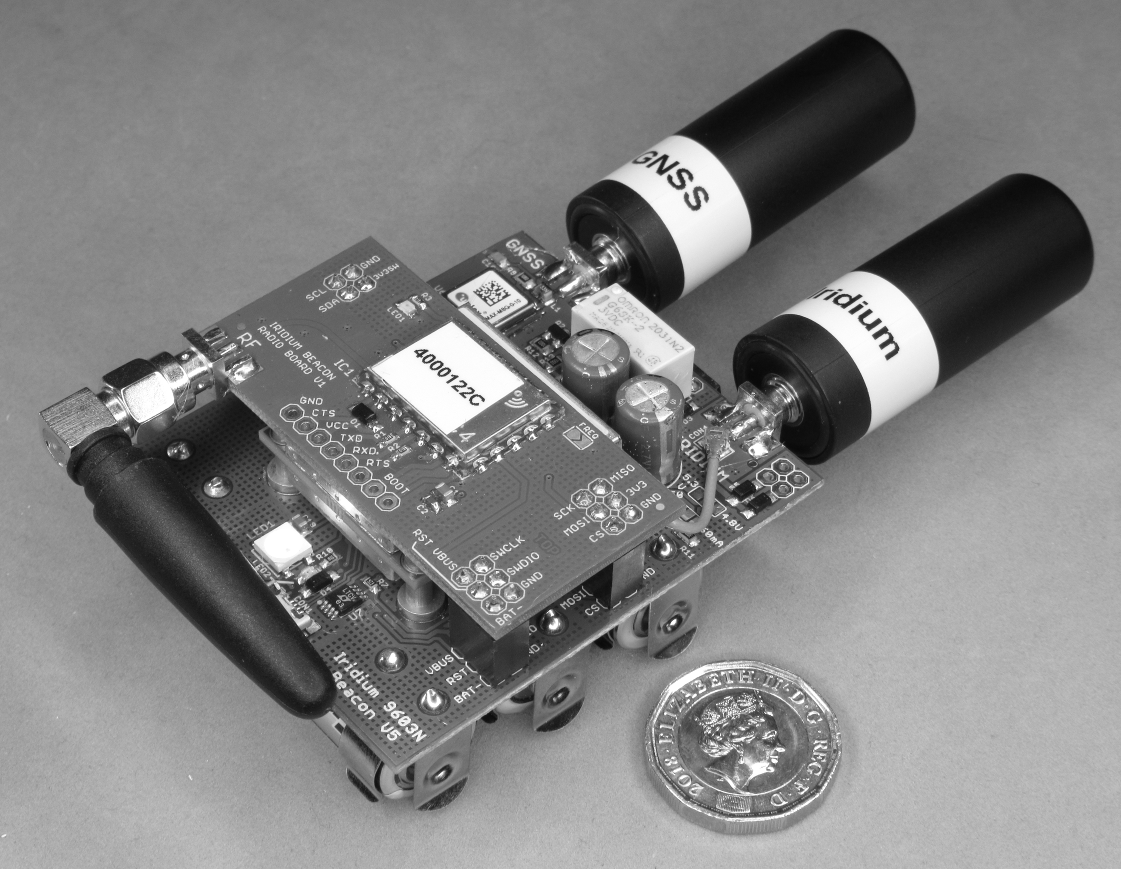}
\caption{\label{fig:beacon} The Iridium 9603 Beacon shown with the optional Radio Board installed. The total weight of the beacon in this configuration is 129\,g.}
\end{figure}

In this section we describe some of the most relevant design considerations of the Iridium 9603 Beacon. Part numbers refer to the schematic and Bill Of Materials (BOM) available on GitHub\textsuperscript{\ref{github}}. The beacon is shown in Fig.~\ref{fig:beacon}.

\subsubsection{SBD Transceiver}
We selected the Iridium\texttrademark\ 9603N short burst data transceiver to provide data communications to and from the beacon as: it is very compact (31.5 $\times$ 29.6 $\times$ 8.10\,mm) and lightweight (11.4\,g); has an operating temperature range of -40\,\textdegree C to +85\,\textdegree C; has modest power requirements; the Iridium constellation of 66 active satellites in six polar orbital planes provide complete global coverage, including the polar regions \cite{b}. It requires a Direct Current (DC) power supply in the range 5.0\,V +/- 0.5\,V with a ripple of less than 40 mV. The average current requirements are: 34 mA when idle; 39mA when receiving data; and 145 mA when transmitting data. The momentary peak current draw during data transmission is 1.3 A; this raises the average current draw during a complete SBD message transfer to 158 mA. The transceiver supports Mobile Originated (MO) (transmitted) messages in binary or text format of up to 340 bytes in length; Mobile Terminated (MT) (received) messages have a maximum length of 270 bytes.

The pins of the 9603N are described using Data Circuit-terminating Equipment (DCE) notation. Although pin 6 is named DF\_S\_TX, the pin is actually an input and needs to be connected to the serial transmit pin of the microcontroller. Likewise pin 7, DF\_S\_RX, is an output and should be connected to serial receive. The transceiver will not function unless the Request To Send and Data Terminal Ready inputs (pins 13 and 14) are pulled low externally.

The 9603N developer's guide describes that it is possible for the modem to be placed in a non-operational state if the requirements for the modem's reset circuit are not met. We have observed this erroneous behavior, albeit rarely, when using an earlier version of the beacon design where the 9603N was powered directly by the supercapacitors. To correct this, power to the 9603N is now switched by a P-channel Field Effect Transistor (FET) (Q4).

Signals are transmitted and received using a Maxtena Inc. M1621HCT-SMA helical SMA antenna. Version 1 of the beacon did use a patch antenna, but we found that using a helical antenna provides superior performance especially after landing horizontally, in forests or on water.

Although there are many international Iridium resellers and SBD service providers, we use 9603N transceivers provided by the UK company Rock Seven Mobile Services Ltd.. Rock7 provide a user-friendly web interface and a gateway service which allows messages from one transceiver to be automatically forwarded to a second. We take advantage of this to allow the Iridium Beacon to be configured and tracked from anywhere, without an Internet connection, using a second beacon acting as a ``base'' and connected via Universal Serial Bus (USB) to a laptop computer or a single board computer such as the Raspberry Pi\textregistered. In section \ref{Python} we describe the Python base software which allows a flight to be tracked and its path visualised using the Google\textregistered{} Static Maps API.

\subsubsection{LTC3225 Supercapacitor Charger}
To allow the beacon to be useful in many asset tracking or remote monitoring applications, not solely for balloon tracking, it has been designed so it can draw power from standard AA batteries, a standard USB port or from lightweight solar panels. Given that a USB port can deliver 5.0\,V power with a maximum current draw of 0.5\,A, and that a lightweight solar panel can deliver substantially less current than that, it is necessary to implement additional power buffering or storage to allow the average and peak current requirements of the 9603N to be met.

We use an Analog Devices Inc. (formerly Linear Technology Corporation) LTC3225EDDB supercapacitor charger (U3) to charge two 2.7\,V supercapacitors, connected in series, to 5.3\,V to provide power for the Iridium 9603N transceiver. The current at which the LTC3225 charges the capacitors is programmable, up to a maximum of 150\,mA. The charging circuit has an efficiency of approximately 50\%, i.e. the total current drawn by the LTC3225 is approximately twice the current delivered to the capacitors when powered by 5.0 V. The charger incorporates an internal change pump and can charge the supercapacitors to 5.3\,V when powered by a supply voltage as low as 2.8\,V. By setting the charging current to the maximum 150 mA, the charger can provide the average current draw of the 9603N transceiver when transmitting data. Therefore 1\,Farad supercapacitors suffice as they only need to hold enough charge to provide the 1.3\,A drawn by the transceiver during its momentary (8.3\,ms) transmissions. This is only possible when the beacon is powered from batteries or USB which can sustain the continuous $>$300\,mA current draw (remembering that the charger efficiency is 50\%).

Lightweight solar panels such as the MPT3.6-150 manufactured by PowerFilm Solar Inc.\ can provide a voltage of 3.6\,V at a current of 100\,mA under full sun and weigh only 2.8\,g each. The beacon can be powered from two of these panels, but it is necessary to limit the maximum current draw to a level which the panels can provide. By replacing the 1\,F supercapacitors with 10\,F ones and reducing the charge current to 60\,mA, it is possible for the panels to provide the 9603N's average current draw when receiving data. The larger 10 F capacitors then provide the additional capacity to meet the average current draw during the complete transmit cycle, not just during the momentary transmissions. Adequate time must be given between transmission attempts to allow the supercapacitors to recharge.

We selected supercapacitors manufactured by Eaton\textregistered{} Bussmann\textregistered{} (C11 and C12). The HV1030-2R7106-R 10 F version has an operating temperature range of -40\,\textdegree C to +65\,\textdegree C, can deliver current pulses of 10 A and weighs 3.2\,g.

\subsubsection{Microcontroller}
We selected the Microchip Technology Inc. (formerly Atmel\textregistered{} Corporation) SAMD21G18 ARM\textregistered{} Cortex\textregistered{} -M0+ microcontroller (U5) as used on the Arduino\textregistered{} Genuino Zero and the Adafruit Feather M0. This device offers 256K bytes of Flash Memory, 32K bytes of Static Random Access Memory (SRAM), a Real Time Clock, and six Serial Communication (SERCOM) interfaces each of which can be configured for Universal Asynchronous Receiver/Transmitter (UART), Inter-Integrated Circuit (I\textsuperscript{2}C) or Serial Peripheral Interface (SPI) operation. We use five of the SERCOMs to provide interfaces for: USB / configuration / diagnostics; GNSS communication; Iridium communication; temperature and pressure sensing via I\textsuperscript{2}C; and, using an additional board\footnote{\label{radio_board}https://github.com/PaulZC/Iridium\_Beacon\_Radio\_Board}, radio communication to trigger the cut-down device(s). The microcontroller's low power modes are used to minimise current consumption between message cycles. The non-volatile Flash memory is used to store the beacon's configuration, the values being preserved even if the power is completely removed.

Although the SAMD21G18 has a built-in power-on reset and brown-out detector circuit, it does not work correctly if the supply voltage rises very slowly. We have found that the device will reset correctly if the power supply ramps up at 0.3 V/s or more, but fails to reset correctly at 0.2 V/s or less. As the solar panel voltage could ramp up very slowly at sunrise, we include a separate reset supervisor (Microchip Technology Inc. MCP111-240) (U1) which holds the microcontroller in reset until the supply rises above 2.4V, ensuring a clean start.

\subsubsection{GNSS Receiver}
Having had problems with the GPS receiver used on version 1 of the beacon, we migrated to the u-blox M8 family of GNSS receivers and have had no difficulties since. The beacon incorporates the MAX-M8Q 72-channel GNSS receiver (U6) which can receive signals from three GNSS systems concurrently. Signals are received using a Maxtena Inc. M1516HCT-P-SMA helical SMA antenna. u-blox do offer versions of the M8 with built-in patch or chip antennas, but we have found that using a separate helical antenna provides superior performance especially after landing horizontally, on ground or on water. The M8's ``dynamic model'' can be optimised for ``Airborne $<$1g'' operation, which is ideal for HAB flight tracking.

\subsubsection{MPL3115A2 Pressure and Temperature Sensor}
\begin{figure}
\centering 
\includegraphics[width=.75\textwidth]{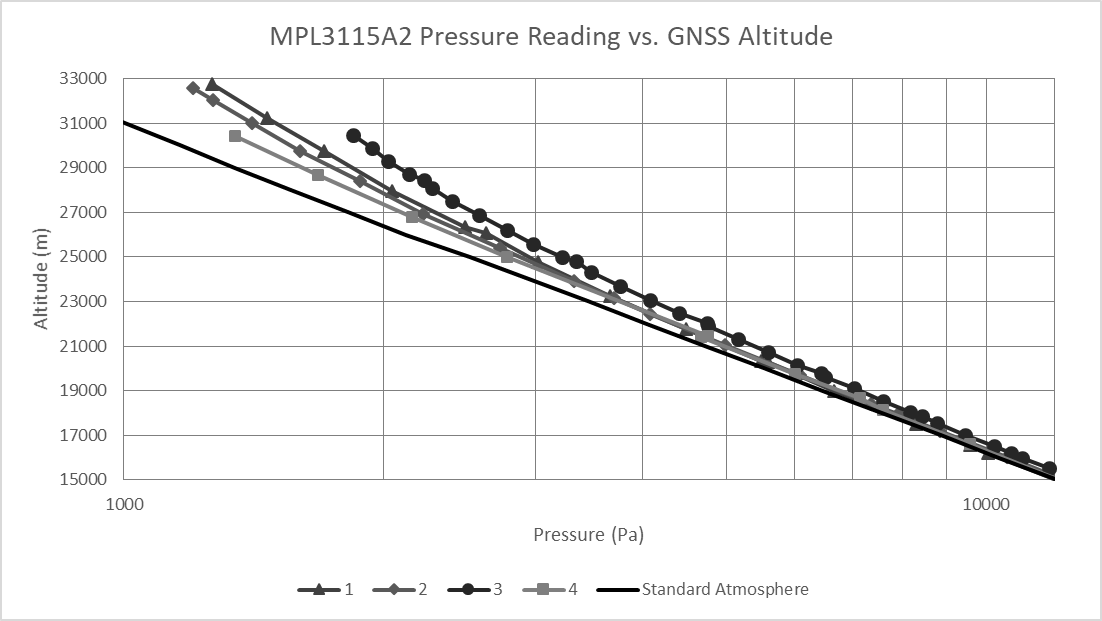}
\caption{\label{fig:MPL3115A2} Scatter plots of the atmospheric pressure readings from four different MPL3115A2 sensors versus the true altitude as measured by the associated MAX-M8Q GNSS receiver. The readings were recorded on four different weather balloon flights. Pressure versus altitude for the U.S. Standard Atmosphere 1976 is included for comparison. The plots illustrate: how the accuracy of the MPL3115A2 pressure readings degrade above 15 km; and the large variation in measured pressure from device to device.}
\end{figure}

The beacon design includes a Freescale Semiconductor Inc. MPL3115A2 combined pressure and temperature sensor (U4) with I\textsuperscript{2}C interface. Its usefulness is limited as flight pressures and temperatures take it well beyond its operational pressure limit of 20 kPa (equivalent to approximately 11,800 m) and operating temperature limit of -40\,\textdegree C. The atmospheric pressure is sensed by the deformation of a Microelectromechanical System (MEMS) membrane. The sensor can estimate altitude directly using a built-in single layer implementation of the barometric formula, however this produces increasing large errors above approximately 15 km. Using a multiple-layer implementation of the barometric formula, implemented in software, improves the conversion of pressure into altitude, however the error is large above 25 km and device to device variations degrade the accuracy further still (see Fig.~\ref{fig:MPL3115A2}). A further complication is that the pressure reading can underflow or wrap-around at extreme altitude (approximately 36 km) and produce pressure readings which are equivalent to being below sea level. Knowing this, the beacon software automatically discards any pressure readings greater than 110 kPa. The reported temperature fails to decrease any lower than typically -47\,\textdegree C even though the true temperature is colder than this.

The pressure and temperature readings are included in the messages transmitted by the beacon but, for the reasons outlined above, they should be used as approximate values only with large error bars.

\subsubsection{Radio Communication}\label{radio_comms}
The first version of the mechanical cut-down device was triggered electrically via a pair of wires connected to the beacon's relay terminals. With the cut-down mounted above the in-line parachute and the beacon mounted below, the wires had to run past the parachute itself. With hindsight, this was a poor choice as the wires could become tangled in the parachute, resulting in the parachute failing to deploy on two flights. Learning from this, communication between the beacon and the cut-down(s) is now via a short range radio link.

We selected the Low Power Radio Solutions (LPRS) easyRadio Integrated Controller (eRIC) to provide the radio link. The eRIC is available in two versions: the eRIC4 provides coverage for the European 433 MHz Industrial, Scientific and Medical (ISM) band including airborne short range devices; the eRIC9 provides equivalent coverage for the 868 MHz  European and 915 MHz USA ISM bands. The eRIC is based on the Texas Instruments CC430F5137 System-on-Chip device. As delivered, it will operate in ``easyRadio'' mode, acting as a UART to packet radio transceiver. However, it can also be reprogrammed with bespoke software which takes advantage of the full easyRadio Operating System (eROS) and the other functionality offered by the CC430F5137.

We have designed a small radio board\textsuperscript{\ref{radio_board}} which carries the eRIC coupled to a stubby antenna and which can be piggy-backed onto the beacon through unused Input / Output (I/O) pins. When MT satellite messages are received by the beacon, they are processed and re-transmitted by the eRIC as required. In this way, it is possible to configure or trigger multiple cut-down devices remotely from anywhere.

\subsubsection{Battery Selection}\label{batteries}
We selected Energizer\textregistered{} Ultimate Lithium batteries as the power source for the beacon. They offer a quoted operating temperature range of -40\,\textdegree{}C to +60\,\textdegree{}C and are relatively lightweight: 7.6\,g, 15\,g and 33.9\,g for the AAA, AA and PP3 versions respectively. We have found that these batteries will continue to provide power at temperatures below -40\,\textdegree{}C, but the variation of battery voltage with temperature needs to be taken into consideration.

\begin{figure}
\centering 
\includegraphics[width=.75\textwidth]{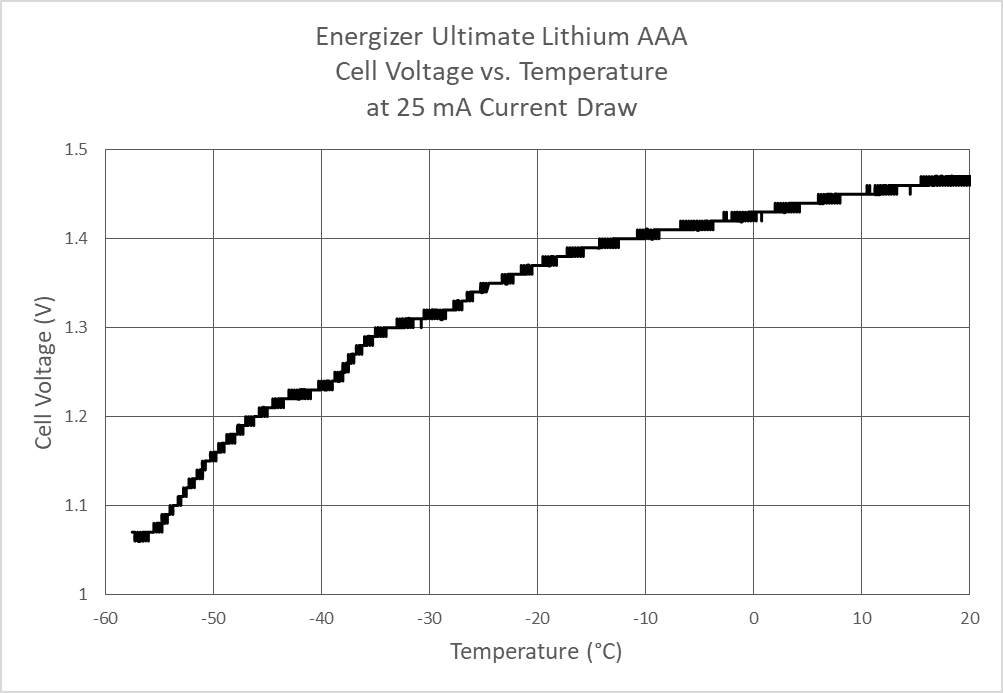}
\caption{\label{fig:energizer} Scatter plot of Energizer Ultimate Lithium AAA cell voltage versus temperature at 25 mA current draw.}
\end{figure}

Fig.~\ref{fig:energizer} illustrates how the voltage of an Energizer Ultimate Lithium AAA cell varies with temperature. At -56\,\textdegree{}C the voltage is approximately 1.07\,V. A new cell at room temperature delivers a voltage of approximately 1.8\,V. We therefore made the decision to power the beacon from three AA cells connected in series as the LTC3225 supercapacitor charger should be able to operate from the 3.2\,V delivered at -56\,\textdegree{}C and can withstand the 5.4\,V delivered by new cells at room temperature.

An SPX3819 low drop-out voltage regulator (U2) delivers the 3.3\,V required by the \\SAMD21G18 and the MAX-M8Q. At -56\,\textdegree{}C the regulator should still deliver approximately 2.9\,V, the drop-out voltage being approximately 0.3\,V at low current. 2.9\,V is just sufficient to power the MAX-M8Q (which had a minimum operating voltage of 2.7\,V). The SAMD21G18 can operate from voltages as low as 1.62\,V.

As the batteries become depleted through use, their voltage drops lower. This results in the beacon shutting down at cold temperatures when the voltage is inadequate to power both the LTC3225 and MAX-M8Q.

\subsubsection{Supply Voltage Monitoring}
The battery or solar panel voltage is monitored by a simple resistive 50\% voltage divider (formed from R3 and R6) which is connected to one of the microcontroller's analogue inputs. When the supply voltage falls below approximately 3.6\,V, the SPX3819 regulator is unable to maintain its 3.3\,V output and the rail drops linearly in line with decreasing supply voltage. As the voltage measured by the analogue input is \textit{relative} to the rail voltage, the supply voltage appears to never drop below 3.4\,V even when the true voltage is substantially lower than this. The beacon includes an Analog Devices Inc. (formerly Linear Technology Corporation) LT1634 1.25\,V micropower voltage reference (U7) which is connected to another of the microcontroller's analogue inputs. As the 3.3\,V rail falls, the voltage produced by the voltage reference appears to \textit{increase} and can be used to correct the supply voltage measurement. In this way, it is possible to determine when the supply voltage has dropped below 2.8\,V at which point the supercapacitor charger is disabled, Iridium communication is no longer possible, and the microcontroller will wait for the voltage to increase before attempting further communication.

\subsubsection{Weight and Weight Reduction}
The beacon weighs 113\,g when powered by Energizer\textregistered{} Ultimate Lithium batteries and equipped with the two Maxtena Inc. helical antennas described above. The addition of the radio board and antenna increases the total weight to 129\,g.

It is possible to reduce the weight of the beacon by having the GNSS receiver and Iridium transceiver share a single Maxtena Inc. M1600HCT-P-SMA antenna, which is tuned for the Iridium, GPS and GLONASS frequency bands. Antenna switching can be performed by a  Skyworks Solutions Inc. AS179-92LF GaAs Radio Frequency (RF) switch. The switch also isolates (protects) the GNSS receiver during Iridium transmissions. The Iridium 9603N Solar Beacon\footnote{https://github.com/PaulZC/Iridium\_9603N\_Solar\_Beacon}, a miniature version of the Iridium Beacon, includes antenna sharing and weighs only 48\,g when powered by two PowerFilm Solar Inc. MPT3.6-150 panels.

\subsection{Software}
The software for the Iridium Beacon is written in C and developed in the Arduino Interactive Development Environment (IDE). The Iridium Beacon is based upon the Adafruit Feather M0 which is in turn based upon the Arduino Zero. The SAMD21G18 processor is programmed with the same bootloader as used by the Adafruit Feather M0, which allows the processor to be configured via USB by the IDE; the compiled software is stored in the processor's Flash memory.

We make use of several open-source software libraries within the beacon code, including Mikal Hart's Iridium SBD library. Written originally for the Rock7 RockBLOCK module, this library provides function call access to the Iridium 9603N transceiver and includes callbacks which prevent the beacon code from blocking during lengthy SBD message cycles.

The beacon software is based upon a switch case statement which governs the sequencing of operation. When the beacon wakes from deep sleep via a Real Time Clock (RTC) alarm interrupt, it: initialises the serial ports; powers up the MAX-M8Q GNSS receiver; waits until the receiver establishes a fix; powers down the GNSS receiver and powers up the LTC3225 supercapacitor charger; waits for the supercapacitors to charge; enables the 9603N transceiver and queues an Iridium message transmission; powers everything down and puts the processor back into deep sleep until the next alarm interrupt. The MO message sent by the beacon contains: GNSS date, time, latitude, longitude, altitude, speed, course (heading), Horizontal Dilution Of Precision (HDOP) and the number of satellites being tracked; atmospheric pressure and temperature from the MPL3115A2; battery voltage and a message sequence number. If an MT message is received during the Iridium message cycle, the message is parsed and processed. Messages denoted for the eRIC radio transceiver are transmitted (three times for redundancy) allowing cut-down devices to be triggered. The message interval, the duration between RTC alarm interrupts, can be configured via MT message. The MO message can be prefixed with the serial number of another Iridium transceiver (``RBDESTINATION'') which the Rock7 gateway will process and \textit{automatically} forward to that transceiver. In this way, a balloon-borne beacon can be tracked from a second beacon acting as a ``base'', connected to a laptop or embedded computer. This makes it possible for a flight to be tracked and controlled from anywhere, without an Internet connection.

\section{Pyrotechnic Cut-Down Device}\label{pyro_device}
\begin{figure}
\centering 
\includegraphics[width=.5\textwidth]{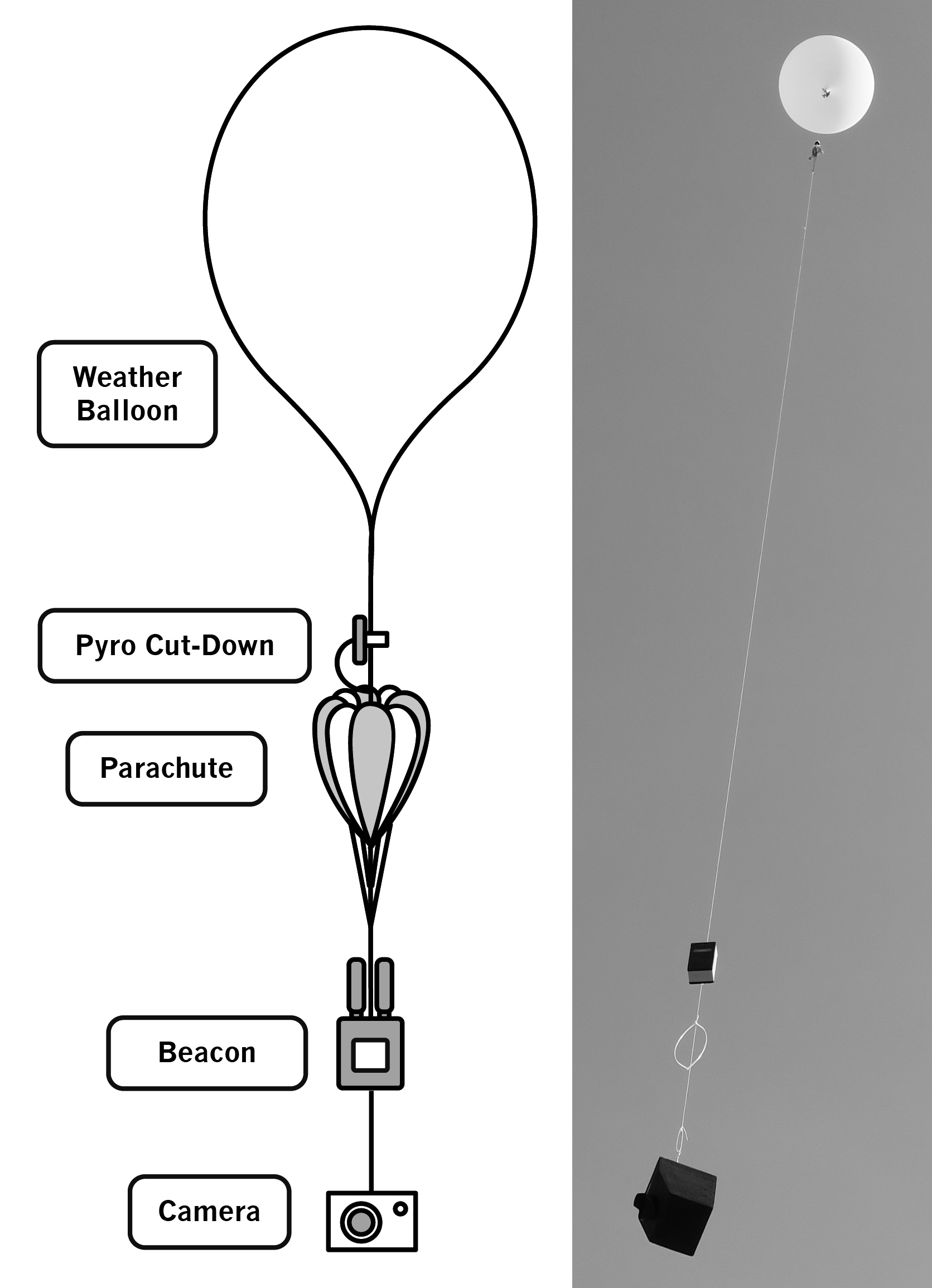}
\caption{\label{fig:balloon} Weather balloon flight train: the illustration shows the relative positions of the balloon, pyrotechnic cut-down, parachute, beacon and camera payload; the photograph shows the increased separation between the balloon \& parachute and the beacon \& camera, the long pendulum length helps the camera to remain stable.}
\end{figure}

On a typical weather or sounding balloon flight, the balloon and payload ascend until they reach an altitude where the reduced atmospheric pressure causes the balloon to expand to its burst diameter. After the balloon has burst, the payload descends on a parachute. Typically the parachute is attached in-line between balloon and payload, the balloon is attached to the top of the parachute canopy and the payload is attached to the parachute's suspension lines (Fig.~\ref{fig:balloon}). Unfortunately, the remains of the burst balloon can cause the parachute to collapse or fail to deploy. A cut-down device overcomes this difficulty by cutting the payload down from the balloon before it bursts. On a dual-balloon flight, two cut-downs may be used: one to release the lift balloon at the desired float altitude; and a second to cut the payload down from the float balloon at the end of the flight. The cut-down device is typically placed on the cord which links the balloon to the top of the parachute canopy and remains attached to the parachute after separation so it can be retrieved. The cut-down must be sufficiently lightweight such that it itself does not cause the parachute to collapse or fail to deploy.

In section \ref{radio_comms} above, we described how the wires linking the first version of the cut-down device to the Iridium Beacon became entangled in the parachute and caused it to fail to deploy. Since then we have refined the design of the cut-down devices such that they can be triggered remotely (via radio) by the beacon.

M., B.\ and T.\ Funk have previously used cut-down devices which use a nichrome wire to melt through nylon cord attaching the balloon to the parachute, but they have found them to be unreliable. It is unclear if it is poor contact between wire and cord that is the main cause of difficulty, or whether the cause is associated with the flight conditions. Based on this experience, we decided not to use hot wire as a means of severing the cord. That said, the design presented in this section can be used to energise a hot wire if required.

The requirements for the cut-down device can therefore be summarised as:
\begin{itemize}
\item To be able to reliably release or sever the cord used to secure the payload or parachute to the balloon(s)
\item To function reliably in the operating environment (atmospheric temperatures and pressures) encountered during flights to typically 32,000 m
\item To be able to reliably determine the altitude during all parts of the flight and to trigger at a pre-programmed altitude limit (if required)
\item To be able to be reliably triggered remotely (via radio) at any point of the flight (if required)
\item The size and mass of the cut-down must be well within the lift capacity of a typical weather balloon and sufficiently lightweight that it can rest on top of the parachute during the descent without causing the parachute to collapse or fail to deploy
\end{itemize}

\subsection{Design Considerations and Component Selection}
\begin{figure}
\centering 
\includegraphics[width=.75\textwidth]{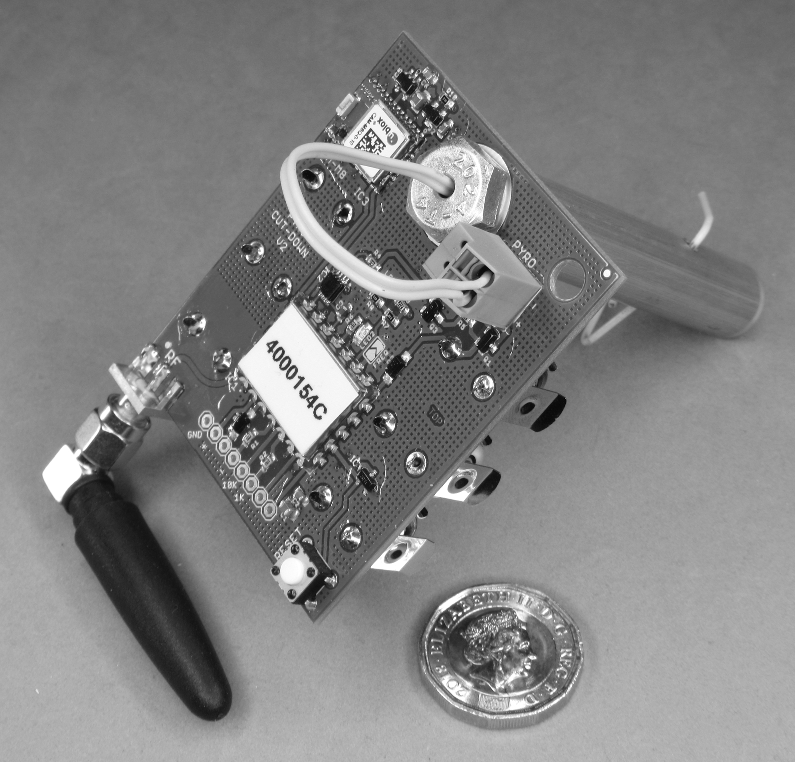}
\caption{\label{fig:pyro} The pyrotechnic cut-down device. The balloon cord would be passed through the cross-drilled hole in the cutter body. A paperclip has been inserted through the hole to prevent the dowel pin from sliding out of position. The pyrotechnic will be triggered and the cord severed at the pre-programmed altitude limit or in response to a radio message. The total weight of the device is 77\,g.}
\end{figure}

To meet these requirements, we have designed and tested a lightweight (77 g) pyrotechnic cut-down device which is able to sever size 4N (2.4 mm) nylon cord. Size 4N cord has a typical breaking strength of 130 kg and exceeds the needs of typical weather balloon flights. In this section we describe some of the most relevant design considerations of the cut-down device. Part numbers refer to the schematic and Bill Of Materials (BOM) available on GitHub\footnote{\label{pyro_github}https://github.com/PaulZC/Pyrotechnic\_Balloon\_Cut-Down}. The device is shown in Fig.~\ref{fig:pyro}.

\subsubsection{Microcontroller and Radio Communication}
We have based the cut-down device on the same LPRS eRIC integrated controller (IC1) used on the Iridium Beacon radio board, described in section \ref{radio_comms}. We take full advantage of the eROS operating system, reprogramming the eRIC such that it will trigger when: (i) it receives its own unique serial number via a radio transmission from the Iridium Beacon; or (ii) reaches a pre-programmed altitude limit. The altitude limit can be set via the I/O pads or via radio message. The eRIC C code is developed and compiled in Texas Instruments' Code Composer Studio.

\subsubsection{GNSS Receiver}
The cut-down is equipped with a u-blox CAM-M8Q GNSS receiver (IC3) which incorporates a small chip antenna and provides accurate altitude data during the flight using the same M8 core as the MAX-M8Q used on the beacon. To save power, the CAM-M8Q is configured to use only GPS satellites and ignores the GLONASS, Galileo and BeiDou constellations. Once the receiver has established a fix, it is put into low power mode to further extend the battery life. The CAM-M8Q can be disabled completely by setting the pre-programmed altitude limit to 99\,km. The NMEA messages produced by the receiver are parsed by the eRIC using Alan Carvalho de Assis' C implementation of Mikal Hart's TinyGPS library.

\subsubsection{7402 NOR gate}
The Input/Output pins on most microcontrollers go into a floating (high impedance) state during reset and can only be configured as outputs once the embedded code starts to execute. We use a 7402 dual-NOR gate (IC2) to ensure that the pyrotechnic is not accidentally triggered during that period. Power for the 7402 is provided by one of the eRIC's I/O pins and the pyrotechnic will only be triggered when another two I/O pins are placed in opposing states (one high, one low). Power to the pyrotechnic is switched by a P-channel FET (Q1).

\subsubsection{Battery Selection}
We selected three Energizer\textregistered{} Ultimate Lithium AAA batteries as the power source for the cut-down, for the same reasons described in section \ref{batteries}. The electric match igniter can be reliably triggered by a single battery, however we apply the combined voltage of all three to the igniter to provide margin for cold conditions when the batteries may be partially depleted.

\subsection{Cord Cutter}
The body of the cord cutter is machined from high strength 7075 T6 aluminium rod. The bore of the cutter is reamed using an over-sized reamer allowing the stainless steel dowel pin to slide easily along it. A small diameter hole at the far end of the cutter body allows the air in front of the dowel pin to escape; the hole is sufficiently small and the cutter body sufficiently strong so that the dowel pin remains captive. The balloon cord is passed through a cross-drilled hole in the cutter body; the cord is severed as the dowel pin passes along the body. The dowel pin is propelled by the gas generated by a small amount (0.1\,g) of smokeless gunpowder; the gunpowder is ignited by a MJG Technologies J-Tek1 igniter normally used in firework displays. The igniter wires pass through a small hole drilled through the center of an aluminium screw; o-rings prevent the igniter from coming into contact with the (conductive) cutter body and restrict any gas escaping through the hole in the screw. The igniter contains a short length of nichrome wire coated with a small amount of pyrotechnic; the pyrotechnic ignites when a current of 1\,A or more is passed through the 1\,$\Omega$ nichrome wire. Debris from the cord can cause the dowel pin to become jammed at the end of the cutter body after use; a parallel pin punch can be passed through the ventilation hole to free the dowel pin leaving the parts ready for cleaning and re-use. New o-rings should be used each time the cutter is re-assembled. The bare ends of the igniter wires should be kept twisted together until they are attached to the circuit board, to prevent the igniter being triggered accidentally by static discharge.

\section{Mechanical Cut-Down Device}
\label{sec:mechanical}

The pyrotechnic cut-down device described in section \ref{pyro_device} has the obvious disadvantage that the igniter and smokeless gunpowder cannot easily be transported by aircraft. To overcome this, we have designed and tested a mechanical version of the cut-down device which utilises a model aircraft servo to open an archery release aid. The archery release aid is perfect for this application since it is designed to release cord which is under heavy load, yet operate with minimal trigger pressure. We use a standard model aircraft metal gear servo which is modified for high altitude operation. Functionally the cut-down operates in the same way as the pyrotechnic version; the servo will open the release aid at a pre-programmed altitude limit or via a radio message from the Iridium Beacon. It does have the disadvantage of being heavier, weighing 115\,g. The mechanical linkage which links the servo and release aid together is integrated into the circuit board design; score lines can be machined into the circuit board to allow the linkage to be snapped off.

\subsection{Design Considerations and Component Selection}
\begin{figure}[htbp]
\centering 
\includegraphics[width=.75\textwidth]{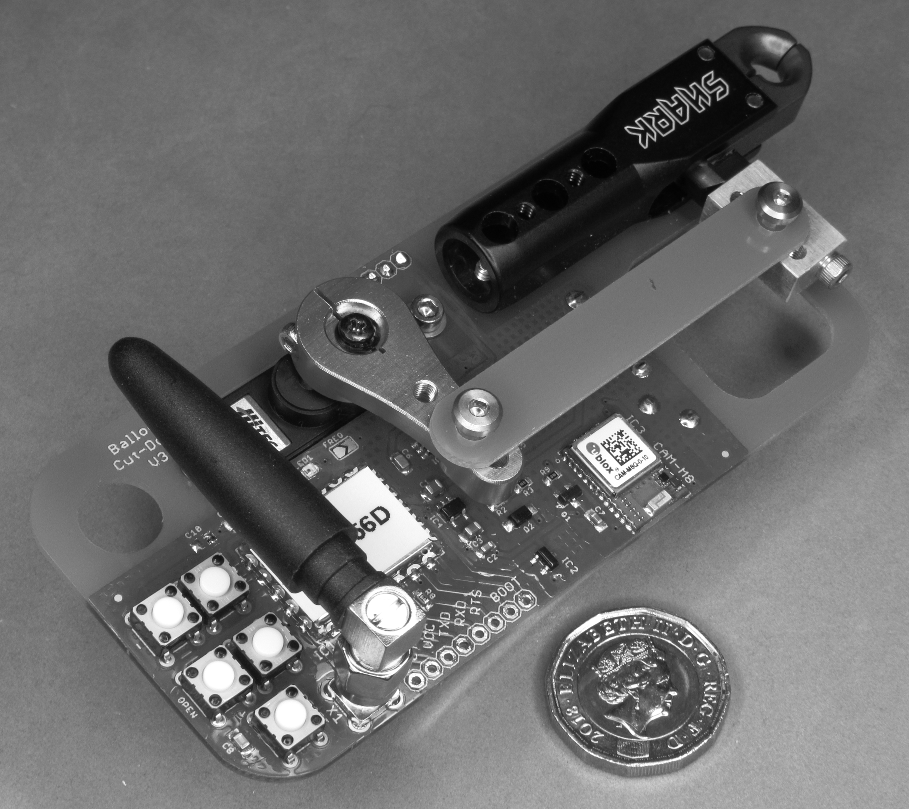}
\caption{\label{fig:mechanical} The mechanical cut-down device, shown in the closed position. The balloon cord would be looped through the archery release aid jaws. The parachute cord would be secured through the circular hole in the circuit board. The release aid will open at the pre-programmed altitude limit or in response to a radio message. The total weight of the device is 115\,g.}
\end{figure}

The mechanical cut-down device has been designed to meet the same requirements as the pyrotechnic version. In this section we describe some of the most relevant design considerations of the cut-down device. Part numbers refer to the schematic and Bill Of Materials (BOM) available on GitHub\footnote{\label{cutdown_github}https://github.com/PaulZC/Balloon\_Cut-Down\_Device}. The device is shown in Fig.~\ref{fig:mechanical}.

\subsubsection{Microcontroller and Radio Communication}
We have based the mechanical cut-down device on the same LPRS eRIC integrated controller (IC1) as used on the pyrotechnic version. We take further advantage of the eROS operating system, reprogramming the eRIC to provide the Pulse Width Modulated signal required to move the servo and storing the settings for the servo open and closed positions in Flash memory. Push switches allow the altitude limit and servo positions to be configured, the altitude limit can also be set via radio message. Like the pyrotechnic version, the cut-down will trigger when: it receives its own unique serial number via a radio transmission from the Iridium Beacon; or reaches the pre-programmed altitude limit.

\subsubsection{Battery Selection}
We selected the Energizer\textregistered{} Ultimate Lithium PP3 battery as the power source for the cut-down. The 9\,V battery voltage is regulated down to 5.0\,V for the servo by a Monolithic Power Systems Inc. MPM3610 compact switch mode regulator (IC5). 3.3\,V for the eRIC is provided by a SPX3819 regulator (IC4). The 3.3\,V PWM signal from the eRIC is converted to 5\,V by a 74AHCT1G125 buffer (IC6).

\subsection{Servo and Archery Release Modification}
To prevent the grease and oil in the servo and archery release from freezing at altitude both are disassembled, the standard grease and oil removed and replaced with Castrol Braycote 601EF. 601EF is a space qualified grease with a minimum operating temperature of -80\,\textdegree{}C. Acrylic conformal coating (lacquer) is applied to the servo's internal circuit board to prevent problems with dewing or icing on descent.

\section{Python Software}\label{Python}
To complement the HAB toolkit hardware, we have developed and tested a suite of Python software tools to aid the tracking and control of HAB flights. The software tools are available in the Python folder of the GitHub repository\textsuperscript{\ref{github}}.

\subsection{Iridium\_Beacon\_GMail\_Downloader\_RockBLOCK}
MO messages transmitted by the Iridium Beacon are processed by an Iridium ground station and forwarded via the Internet to Rock7's server. The Rock7 Operations web interface is used to configure where the messages are forwarded to next. Forwarding via email and HTTP Post are both possible. As the use of HTTP Post requires the use of a server to receive the posted messages, the software tools described here all use message delivery via email.

Iridium\_Beacon\_GMail\_Downloader\_RockBLOCK.py uses the Google\textregistered{} GMail\textregistered{} Application Programming Interface (API) to check for the arrival of new MO Iridium Beacon messages delivered via email. Periodically, the software checks for the arrival of new, unread SBD messages with an attachment in GMail. When a new message is found, the .SBD message attachment is downloaded to the local computer, the message tagged as read and moved out of the GMail inbox to an archive folder. In this way, the contents of the MO messages from multiple beacons can be downloaded automatically to the local computer within seconds of being transmitted.

\subsection{Iridium\_Beacon\_Mapper\_RockBLOCK}
Iridium\_Beacon\_Mapper\_RockBLOCK.py checks periodically for the appearance of new MO message attachments (the .SBD attachments downloaded by Iridium\_Beacon\_GMail\_Downloader\_\\RockBLOCK.py). When a new attachment is found, it is parsed, the beacon location extracted and its position displayed in a Python Tool Command Language (TCL) Graphical User Interface (GUI) Toolkit (Tk).  The beacon's latitude, longitude, altitude and other flight data are displayed together with the beacon's location overlaid on a map image from the Google\textregistered{} Static Map API. As the flight progresses, the path the beacon has followed is displayed as a coloured line overlaid on the static map image. The paths of up to eight beacons can be tracked and displayed simultaneously. The map zoom level and center position can be adjusted by clicking the Tk buttons or by clicking within the map image. A screenshot of the mapper is shown in Fig.~\ref{fig:flight_19-2-10}.

\subsection{Iridium\_Beacon\_Base}
Iridium\_Beacon\_Base.py operates in a very similar way to Iridium\_Beacon\_Mapper\_\\RockBLOCK.py except that the software expects to be connected to an Iridium Beacon, acting as a ``Base'', via USB. Instead of messages from the beacon(s) arriving as email attachments, message delivery is via the Iridium network directly to the base.

A pull-down menu in the TCL Tk software allows a MO message to be sent to a beacon by the base; the RockBLOCK serial number of the beacon is included as the message prefix, the Rock7 gateway then automatically forwards the message to the beacon. If the message contains the ``RBDESTINATION'' of the base, then all subsequent messages sent by the beacon will be prefixed with the base's serial number and will be automatically be forwarded to the base by the Rock7 gateway from then on. Although the email messages can still be still generated, they are redundant in this case.

Messages from multiple beacons can only be downloaded by the base one at a time; each SBD message cycle only allows one MT message to be received and one MO message to be transmitted. If messages are being sent by two beacons every 10 minutes, the base will need to check for the arrival of new messages every five minutes in order to keep up. Message costs can become non-trival as message credits are used each time a beacon sends an MO message \textit{and} each time a base receives an MT message -- i.e. the message costs are double compared to using email delivery as described above. However, to be clear, message delivery directly from beacon to base allows the beacon to be tracked, configured and controlled by the base from anywhere, without requiring an Internet connection.

Google\_Static\_Maps\_Tiler.py can be used to download Google Static Map images or tiles, while an Internet connection is available, prior to the balloon flight. The position of the beacon(s) can then be displayed overlaid on these map tiles, off-line, during the flight.

\subsection{Iridium\_Beacon\_Stitcher\_RockBLOCK}
Iridium\_Beacon\_Stitcher\_RockBLOCK.py is a simple tool which will concatenate all received .SBD message attachments into a single Comma Separated Value (CSV) file. The data in this file can be processed with Python or Microsoft Excel to display (e.g.) the balloon ascent rate.

Iridium\_Beacon\_CSV\_DateTime.py will change the beacon message date and time from YYYYMMDDHHMMSS format into a more user-friendly DD/MM/YYYY,HH:MM:SS format.

Iridium\_Beacon\_DateTime\_CSV\_to\_KML\_RockBLOCK.py will convert the concatenated flight data from CSV format into the Keyhole Markup Language (KML) format used by Google Earth. The flight path can then be visualised in three dimensions, overlaid on satellite imagery as shown in Fig.~\ref{fig:flight_18-11-24}.

\section{Flight Test Results}
\label{sec:flighttests}

\begin{figure}[htbp]
\centering 
\includegraphics[width=\textwidth]{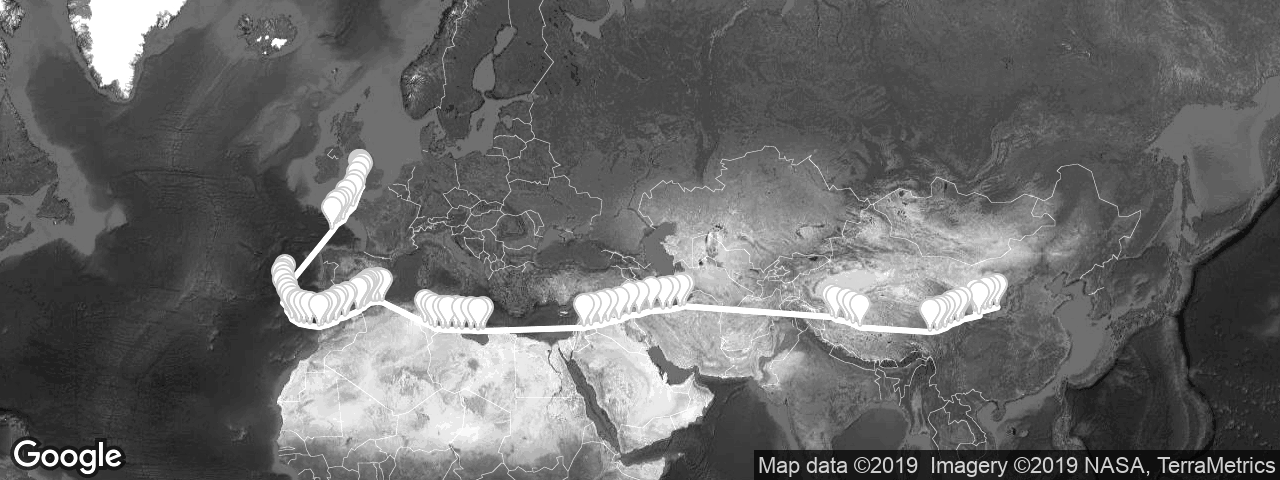}
\caption{\label{fig:UBSEDS22} The flight path of the UBSEDS22 superpressure balloon, March 2017. During each valid telemetry burst, the Iridium Network determined the beacon's location by tracking its position in the satellite's service beams. Each location has a typical circular error probable of approximately 5km and is represented by a marker. The gaps between the markers show where the nighttime temperature caused the beacon to stop transmitting; the temperature above Morocco on the second night was sufficiently high to allow the beacon to continue transmitting.}
\end{figure}

Clark published version~1 of the HAB toolkit design of the Iridium 9603 Beacon in December 2016 on GitHub\footnote{\label{github}https://github.com/PaulZC/Iridium\_9603\_Beacon}. Shortly afterwards the design received welcome publicity via a blog post on the website Hackaday.com\footnote{https://hackaday.com/2016/12/19/a-beacon-suitable-for-tracking-santas-sleigh/}. Meadows and colleagues from the University of Bristol Students for the Exploration and Development of Space (UBSEDS) made contact with Clark and flew version~1 of the Beacon on a hand-made superpressure balloon in March 2017. Although this early GPS receiver module failed to provide position information at altitudes above 10\,km, messages were received from the Beacon for seven days as it was carried from the UK, along the Mediterranean, finally reaching China before the batteries became depleted (see Fig.~\ref{fig:UBSEDS22}). 
M.\ Funk made contact with Clark in October 2017 and in an ongoing collaboration Clark continues to develop and enhance the toolkit while M., B.\ and T.\ Funk use it to track and terminate flights as often as time and weather permit. 

The flight of weather balloons within United Kingdom airspace falls under The Air Navigation Order (ANO) 2016. The Order defines a ``small balloon'' as ``a balloon of not more than two metres in any linear dimension at any stage of its flight, including any basket or other equipment attached to the balloon''. The flight of a single small balloon does not need to be notified to the Civil Aviation Authority (CAA) as this is covered by an exception from application of provisions of the Order. As the majority of weather balloon flights exceed the two metre linear dimension, particularly at altitude, it is necessary to apply for an exemption from the provisions of the Order from the CAA at least 28\,days in advance of the proposed flight. A launch window of a few days can be requested. While this does not prevent pre-planned weather balloon flights, it does make it difficult to take advantage of favorable weather conditions when they occur.

In Switzerland, Article 16 of the Ordonnance du DETEC sur les a{\'e}ronefs de cat{\'e}gories sp{\'e}ciales (OACS) (24\textsuperscript{th} November 1994) prohibits the launch of free balloons: if they are inflated with flammable gas; if their payload is greater than 2\,kg; if their total capacity is greater than 30\,m\textsuperscript{3}. Additional restrictions apply for free balloons launched less than 5\,km from civilian or military airports. Article 20 requires the balloon operator to have public liability insurance of at least 1 million francs. M., B.\ and T.\ Funk are able to launch helium weather balloon flights from the Geneva area, in compliance with the OACS requirements, as often as time and weather conditions permit. In less than 12\,months, 23 flights have been tracked using version~5 of the Iridium 9603 Beacon. The payloads from all but one of those flights have been successfully recovered, including those which have landed on water or in forests. The services of helicopter pilots, boat owners and tree surgeons have been required more than once. Sections \ref{flightdisko} to \ref{vienna} provide more details about three of the flights which allow us to demonstrate the technology readiness level of the toolkit. The one payload which has not yet been recovered is described in section \ref{vienna}.

\subsection{Flight 13/03/17: Bristol, UK}
An exemption to the ANO was obtained by Meadows for a four day window in March 2017. The intention to launch on the third day of this window (13\textsuperscript{th} March) was confirmed with the CAA on the 9\textsuperscript{th} March, and a Notice to Airmen (NOTAM) was subsequently issued. In compliance with the exemption, the UBSEDS team launched the balloon at 05:20 local time from Bristol. In addition to the version 1 Beacon, a small solar-powered tracker of Meadows' design was also present on the flight. This tracker, which did not use the Iridium network, returned data intermittently through the flight\footnote{\label{ubseds22_website}https://www.bristol-seds.co.uk/hab/flight/2017/03/13/ubseds22.html}. The last datapoint was received from 12km above Lake Lietvesi, Finland on 21\textsuperscript{st} May 2017 at 19:21. At this time the sun was below 0\textdegree{} elevation (but still above the horizon at this altitude) and the temperature of the external thermistor was reported as -51\,\textdegree{}C.

\subsection{Flight 05/08/18: Disko, Greenland}\label{flightdisko}
During a vacation in Greenland, M., B.\ and T.\ Funk were able to launch and recover five balloon flights. One flight was launched from a large boat, landed intentionally in the sea and was recovered by the same boat. A second flight was launched from land but also landed in the sea and was again recovered by boat. The beacons on both flights continued to transmit while floating on the sea, encased in extruded polystyrene foam.

\begin{figure}
\centering 
\includegraphics[width=0.8\textwidth]{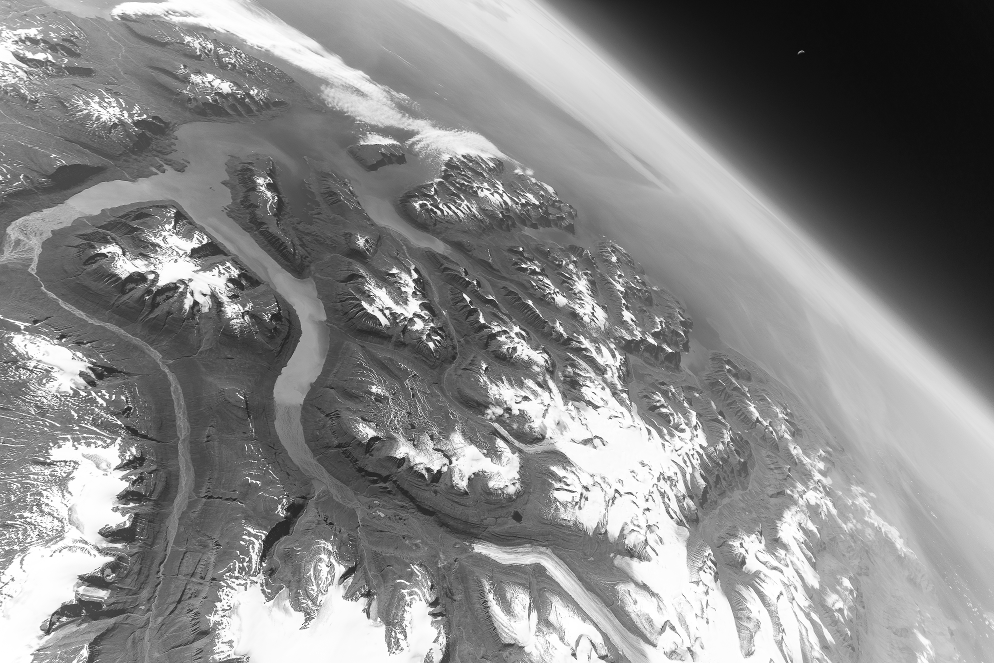}
\caption{\label{fig:disko} Image taken from an altitude of approximately 23,000\,m above Disko, Greenland, on 5\textsuperscript{th} August 2018. The camera is pointing west; the crescent Moon is just visible above the horizon. This image was stored on the SD card which was thrown out of the camera body when the payload crash landed, the parachute having failed to deploy. \textit{Image credits: M., B.\ and T.\ Funk}}
\end{figure}

The flight launched from the island of Disko on the 5\textsuperscript{th} August 2018 ended differently. An attempt was made to use the first version of the mechanical cut-down device; the cut-down was attached to the cord between the balloon and parachute canopy and linked by wires to the beacon attached below the parachute's suspension lines. The flight ascended to 26,000 m where an attempt to open the cut-down via Iridium message was made. The flight continued to ascend, the cut-down having failed to open. The balloon burst at an altitude of approximately 32,000\,m and the payload fell rapidly, reaching speeds of 70\,m/s in the first part of the descent. When the payload hit the ground, it was still traveling in excess of 20\,m/s. The Iridium Beacon sent four transmissions during the 12\,minute descent and continued to transmit after the crash landing. The payload landed in a mountainous area and was recovered eight days later when M., B.\ and T.\ Funk returned to the site by helicopter with Nick Nielsen who abseiled down the rock face to find and recover the payload. The camera flown on the flight had been destroyed in the crash; the SD card had been thrown out of the camera body on impact but fortunately NN was able to recover it. The image shown in Fig.~\ref{fig:disko} was one of many taken during the flight, found preserved on the SD card. It was found that the wires linking the beacon to the cut-down had become tangled in the parachute, preventing it from opening. Learning from this, the cut-down was promptly re-designed and the radio link introduced.

\subsection{Flight 24/11/18: Geneva to Lake Constance}
For the flight launched from Geneva on the 24\textsuperscript{th} November 2018, the balloon had deliberately been filled with less helium than usual to limit the ascent rate to 2\,m/s. Launched at 16:45 local time, the flight was carried ENE parallel to the Alps. Contact was lost with the flight at 20:30 local time at an altitude of 16,200\,m when the batteries became too cold and the voltage too low for the supercapacitors to charge. The reported temperature from the MPL3115A2 sensor had reached -47\,\textdegree{}C 40 minutes earlier, the true temperature was colder than this. The reported battery voltage had fallen to 3.3\,V when contact was lost.

Messages from the flight started to be received again over four hours later. The payload had landed in Lake Constance, had warmed up when in contact with the water and resumed transmissions. The payload was recovered the following afternoon by boat.

\begin{figure}
\centering 
\includegraphics[width=1.0\textwidth]{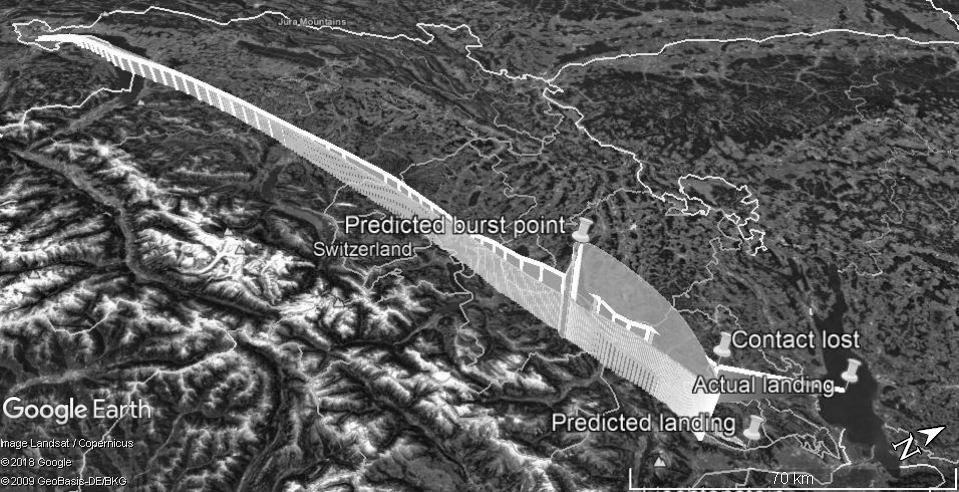}
\caption{\label{fig:flight_18-11-24} KML data showing the actual and predicted flight path of the flight on 24\textsuperscript{th} November 2018. Contact was lost with the balloon at 16,200 m when the nighttime temperature fell. Contact was reestablished when the payload warmed up after landing in Lake Constance.}
\end{figure}

It was difficult to explain how the flight had come to land in Lake Constance given that the previous message was from a location 30\,km to the SW. The answer became clear when the flight path was retrospectively predicted using wind data from the National Centers for Environmental Prediction (NCEP) Global Forecast System (GFS). At 18,000\,m the wind direction reversed, carrying the flight WSW. If the balloon burst between 33,000\,m and 34,000\,m, the flight would indeed have landed only a little farther to the NE from where contact was lost. The actual and predicted flight data is shown in Fig.~\ref{fig:flight_18-11-24}.

We regard the flight as a complete success, the beacon having: shut down as expected when it became too cold; re-started transmissions again after landing in the Lake; travelled a great circle distance of 292\,km.

The calculator written to calculate the required volume of helium to produce the 2 m/s ascent rate is available on GitHub\footnote{\label{calculator}https://github.com/PaulZC/Balloon\_Calculator}. The calculator uses Cardano's method to solve the cubic equation for the balloon diameter, based on the drag and free lift required to produce the desired ascent rate.

\subsection{Flight 10/02/19: Geneva to Vienna}\label{vienna}
\begin{figure}
\centering 
\includegraphics[width=1.0\textwidth]{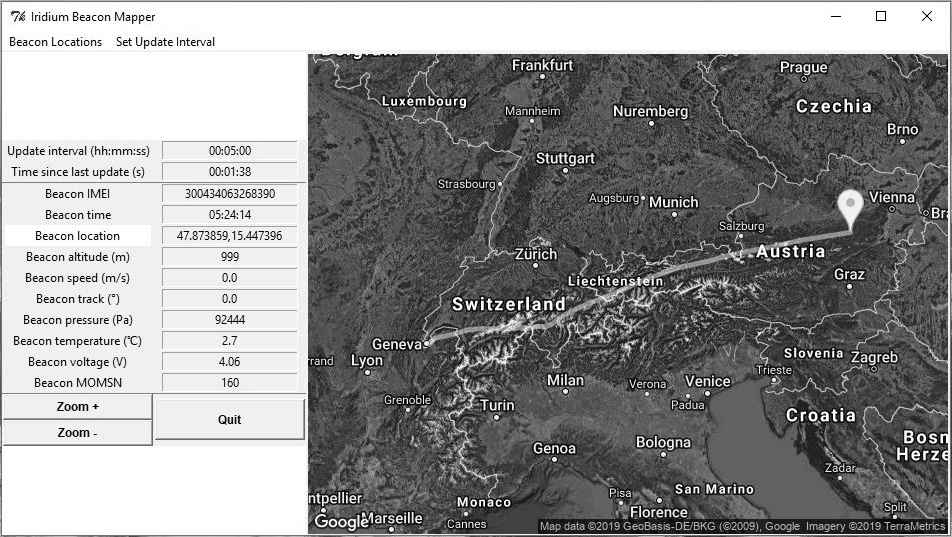}
\caption{\label{fig:flight_19-2-10} The flight path of the dual-balloon flight launched on 10\textsuperscript{th} February 2019, displayed by Iridium\_Beacon\_Mapper\_RockBLOCK.py. The lift balloon was released at 24,000 m above Valais. The flight floated at 29,000 m until the float balloon burst. The payload landed in forest SW of Vienna.}
\end{figure}

The flight launched from Geneva on the 10\textsuperscript{th} February 2019 is the most recent of the 23 flights and spans the largest great circle distance of this series to date. The flight was launched with both lift and float balloons, both balloon cords were equipped with pyrotechnic cut-downs. The lift balloon was released after 1 hour 20\,minutes by pyrotechnic cut-down via radio at 24,000\,m above Valais. The flight then floated at approximately 29,000\,m for a further 4\,hours 40\,minutes at which point the float balloon burst. The flight landed in forest SW of Vienna having traveled a great circle distance of 727\,km. The transmission interval was changed to 12\,hours and at the time of writing the beacon continues to transmit, the batteries having lasted 43\,days so far. The flight path displayed by the Python mapping software Iridium\_Beacon\_Mapper\_RockBLOCK.py is shown in Fig.~\ref{fig:flight_19-2-10}. The calculator written to calculate the neck lift of the lift and float balloons is available on GitHub\textsuperscript{\ref{calculator}}.

\section{Conclusions}

We have presented an open source toolkit of electronic devices and software which have been used to successfully track and recover 23 high altitude balloon flights to date. Together with the additional resources available on GitHub, we have demonstrated that the designs are technologically ready for use as-is on similar projects and have described the designs in sufficient detail that they could be modified for use in many other asset tracking or remote monitoring projects. Brown has used the same toolkit to confirm the location of Unmanned Aerial Vehicles during test flights to aid calibration of the High Energy Stereoscopic System (HESS) for example. Our intention is to integrate the beacon and cut-down designs to deliver the parachute-based terabyte data retrieval system for missions like \textsc{SuperBIT} in the near future.

We consider the Technology Readiness Level (TRL) of the Iridium Beacon to be TRL 9, given that the actual technology has been qualified through 23 successful mission operations. The pyrotechnic cut-down device has been used to successfully control and terminate three flights to date; the mechanical cut-down device has only been used once successfully to date. We therefore describe the TRL of both as 7; the technology prototype has been demonstrated in an operational environment.

\acknowledgments

We are grateful to Roger Smith, Russel Smith and all the {\sc SuperBIT} team for advice and discussions. 
We thank Cl\'emence Cambourian, Romann Heurlin, Josh English, April Higgin and Agaraj Duara for assistance with hardware.
We gratefully acknowledge the time and effort REM's fellow students devoted to preparing the UBSEDS22 superpressure balloon fight.

This project would not have been possible without the open source designs, software and tutorials provided by: Arduino (the Arduino IDE, SAMD board library, RTCZero library); Adafruit (the design of the Feather M0 Adalogger, SAMD board library, SERCOM examples, NeoPixel library); Mikal Hart (the Iridium SBD library, TinyGPS library, PString); Sparkfun (the design of the MPL3115A2 breakout board); Cave Moa (SimpleSleepUSB); Martin L (SERCOM examples); Cristian Maglie (FlashStorage library); LPRS (eROS examples); Alan Carvalho de Assis (C implementation of Mikal Hart's TinyGPS); Steve Randall and Adam Grieg (the Cambridge University Spaceflight (CUSF) balloon flight predictor); Panu Lahtinen (pyBalloon flight predictor). 

Funding was provided by the UK agency STFC, through an Impact Acceleration Award to Durham University, plus grants ST/P000541/1 and ST/N001494/1. 
RJM is supported by the Royal Society.


\end{document}